\documentclass[%
    prl,
    reprint,
    superscriptaddress,
    amsmath,amssymb,
    aps,
]{revtex4-2}

\usepackage{graphicx}
\usepackage{hyperref}
\usepackage{dcolumn}
\usepackage{bm}
\usepackage{physics}
\usepackage{color}

\begin{document}

\title{Observation of Nuclear-wavepacket Interference \\ in Ultrafast Inter-atomic Energy Transfer}%

\author{Meng Han}
\affiliation{ Laboratorium f\"{u}r Physikalische Chemie, ETH Z\"{u}rich, 8093 Z\"{u}rich, Switzerland}
\author{Jacqueline Fedyk}
\affiliation{Theoretische Chemie, Physikalisch-Chemisches Institut, Universität Heidelberg, Heidelberg, Germany}
\author{Jia-Bao Ji}
\affiliation{ Laboratorium f\"{u}r Physikalische Chemie, ETH Z\"{u}rich, 8093 Z\"{u}rich, Switzerland.}
\author{Victor Despré}
\affiliation{Theoretische Chemie, Physikalisch-Chemisches Institut, Universität Heidelberg, Heidelberg, Germany}
\affiliation{Institut Lumière Matière, UMR5306 - UCBL and CNRS, Lyon, France}
\author{Alexander I. Kuleff}
\email{alexander.kuleff@pci.uni-heidelberg.de}
\affiliation{Theoretische Chemie, Physikalisch-Chemisches Institut, Universität Heidelberg, Heidelberg, Germany}
\author{Hans Jakob Wörner}
\email{hwoerner@ethz.ch}
\affiliation{ Laboratorium f\"{u}r Physikalische Chemie, ETH Z\"{u}rich, 8093 Z\"{u}rich, Switzerland}

\date{\today}

\begin{abstract}
We report the experimental observation of quantum interference in the nuclear wave-packet dynamics driving ultrafast excitation-energy transfer in argon dimers below the threshold of interatomic Coulombic decay (ICD). Using time-resolved photoion-photoion coincidence spectroscopy and quantum dynamics simulations, we reveal that the electronic relaxation dynamics of the inner-valence $3s$ hole on one atom leading to a $4s$ or $4p$ excitation on the other one is influenced by nuclear quantum dynamics in the initial state, giving rise to a deep, periodic modulation on the kinetic-energy-release (KER) spectra of the coincident Ar$^+$-Ar$^+$ ion pairs. Moreover, the time-resolved KER spectra show characteristic fingerprints of quantum interference effects during the energy-transfer process. Our findings pave the way to elucidating quantum-interference effects in ultrafast charge- and energy-transfer dynamics in more complex systems.
\end{abstract}

\maketitle
Atoms and molecules can absorb photons of energy exceeding the binding energies of their core or inner-valence electrons, forming highly excited cationic states that can relax through fluorescence or Auger decay. An isolated atom or molecule with an inner-valence vacancy will in general not have enough excess energy to undergo Auger decay, and will relax by fluorescence or, in the case of molecules, also by a coupling to nuclear dynamics, which both are relatively slow processes. In 1997, Cederbaum \textit{et al.}~\cite{cederbaum1997giant} proposed a new mechanism of ultrafast electronic decay where the environment plays a crucial role, termed inter-molecular(atomic) Coulombic decay (ICD), where an inner-valence hole can transfer its energy to a neighboring species which subsequently releases
the excess energy by emitting an electron from its outer-valence shell. After both theoretical \cite{santra2000interatomic,santra2002non,scheit2004on,averbukh2004mechanism,kuleff2007tracing,demekhin2009interatomic,gokhberg2014site,stumpf2016role} and experimental  \cite{marburger2003experimental,jahnke2004experimental,morishita2006experimental,sisourat2010ultralong,ren2018experimental,ren2022ultrafast} studies over more than two decades, ICD was found in a large variety of weakly bound systems, including liquid water \cite{zhang2022intermolecular}, or literally {\it everywhere}~\cite{Sakai2011electron} (for a recent review, see Ref.~\cite{Jahnke2020interatomic}). 

Interestingly, for all homonuclear noble-gas dimers heavier than neon, the ICD channel of the inner-valence hole state is energetically closed. For example, in Ar atoms the $3s$ ionization energy is 29.24~eV which is below the threshold for ICD at 31.52~eV \cite{lablanquie2007appearance}. However, a recent pioneering study on the argon dimer \cite{mizuno2017time} revealed that although the relaxation energy of the $3s$ hole on one atom is not enough to ionize the neighboring atom, it is sufficient to excite an electron of the latter to Rydberg states, creating a dissociative Ar$^+$-Ar$^*$ system. This process can be called pre-ICD or frustrated ICD \cite{lablanquie2007appearance,thissen1998photoionization}.

\begin{figure}
\begin{center}
\includegraphics[width=8.6cm]{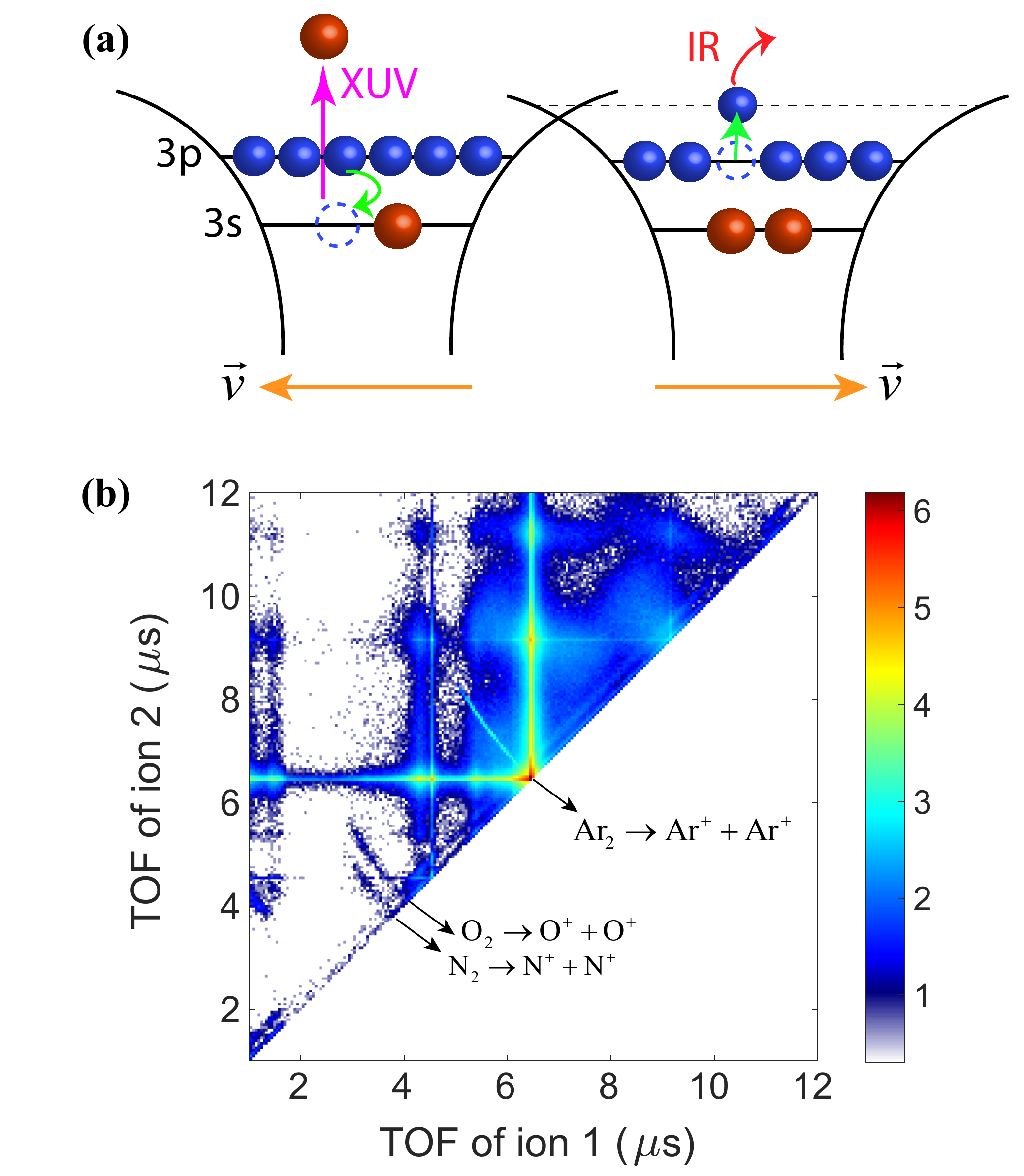}
\end{center}
\caption{(\textbf{a}) Illustration of the pump-probe principle for measuring the frustrated-ICD process of the $3s$-hole in argon dimer. The XUV-pump pulse removes a $3s$ electron from one of the atoms of the dimer (magenta arrow) initiating a frustrated-ICD process, in which a $3p$-electron fills the $3s$-vacancy transferring the gained energy to the other argon where an electron is excited to a $4s$ or $4p$ orbital (green arrows). The excited electron is then ionized by a delayed IR pulse (red arrow) creating a dication that undergoes a Coulomb explosion. (\textbf{b}) Measured photoion-photoion coincidence (PIPICO) spectrum, where the XUV-IR delay was integrated from zero to 2~ps.}
\label{figure1}
\end{figure}

Here, we report the discovery of quantum-interference effects in the nuclear dynamics that mediate the ultrafast inter-atomic energy transfer process (or frustrated ICD) in argon dimers. We find that this quantum interference strongly modulates the energy-transfer probability as a function of time elapsed since the excitation process. As illustrated in Fig.~\ref{figure1}(a), an extreme-ultraviolet (XUV) pulse (magenta) creates a $3s$ hole in one of the Ar atoms of the dimer, followed by the ultrafast transfer (green) of excitation energy to the neighboring argon. A delayed infrared (IR) pulse (red) can then ionize the excited electron from the neighboring atom, which initiates a Coulomb explosion (orange) of the two cations. By monitoring the kinetic-energy-release (KER) spectrum of coincident Ar$^+$-Ar$^+$ ions, one can directly reconstruct the nuclear dynamics in this ultrafast inter-atomic energy transfer process. The equilibrium internuclear distance of the neutral argon dimer is 3.8~\AA, corresponding to a KER of around 3.8~eV. Because the final state of frustrated ICD, Ar$^{+}$-Ar$^*(3p^{-1}nl)$, is a dissociative state, the corresponding KER signal is expected to appear at and below 3.8~eV. Note that the frustrated ICD itself (i.e. in the XUV-only case) does not contribute to the Coulomb explosion signal, as opposed to the ICD process of satellite states Ar$^{+}(3p^{-2}nl)$-Ar created by shake-up ionization \cite{miteva2014interatomic,ren2016direct,rist2017comprehensive}. In an IR field, the satellite states can be ionized and then decay through the radiative charge transfer (RCT) process Ar$^{2+}(3p^{-2})$-Ar$\rightarrow$Ar$^{+}$-Ar$^{+}$ \cite{saito2007evidence,johnsen1978measurements} when the internuclear distance is shrinking, corresponding to a KER peak centered at 5.3~eV. In our experiments, we did not observe the shake-up ICD channel due to its low cross section in our energy regime.

Experimentally, the XUV pulses are obtained from high-order harmonic generation (HHG) by focusing a 30-fs IR pulse into a 3-mm-long gas cell filled with 40~mbar of krypton, with the discrete odd harmonics ranging from 17~eV to 38~eV after spectral filtering through a 200-nm-thick aluminum foil. The XUV-IR cross-correlation signal is measured to be around 40~fs. The time zero is accurately calibrated by the delay-dependent Ar$^{2+}$ yield (not shown) from monomers due to the laser-enabled Auger decay \cite{ranitovic2011laser}. The intensity of the IR pulse is controlled at around $6\times10^{12}$~W/cm$^2$, which is too weak to ionize neutral argon atoms on its own. The XUV and IR pulses are both focused into the reaction center of the COLd Target Recoil Ion Momentum Spectroscopy (COLTRIMS) reaction microscope \cite{dorner00a,ullrich03a}, where the supersonic gas-jet of argon atoms (backing pressure of 1~bar) is delivered by a small nozzle with an opening-hole diameter of 30~$\mu$m and passed through two conical skimmers located 10~mm and 30~mm downstream with a diameter of 0.2~mm and 1~mm, respectively. The fraction of dimers to monomers is measured to be about 3~$\%$. In the COLTRIMS spectrometer, a strong static electric field ($\sim 7.7785$~V/cm) is applied to collect the ions with high kinetic energies and thus the electron detector is not used.

The measured delay-averaged photoion-photoion coincidence (PIPICO) spectrum is reported in Fig.~\ref{figure1}(b), where the diagonal-like features correspond to the ion pairs in Coulomb explosion channels. 
The signal converging to the time-of-flight (TOF) of 6.4 $\mu$s is the reaction pathway Ar$_2$ $\stackrel{\rm{XUV+IR}}{\longrightarrow}$ Ar$^{+}$-Ar$^{+}$, which we are interested in. Using a three-dimensional recoil momentum sphere as a confinement condition in off-line data analysis, we can only pick up the events in this channel and rule out the contribution from residual N$_2$ and O$_2$ gases in the reaction chamber.

\begin{figure}
\begin{center}
\includegraphics[width=8.6cm]{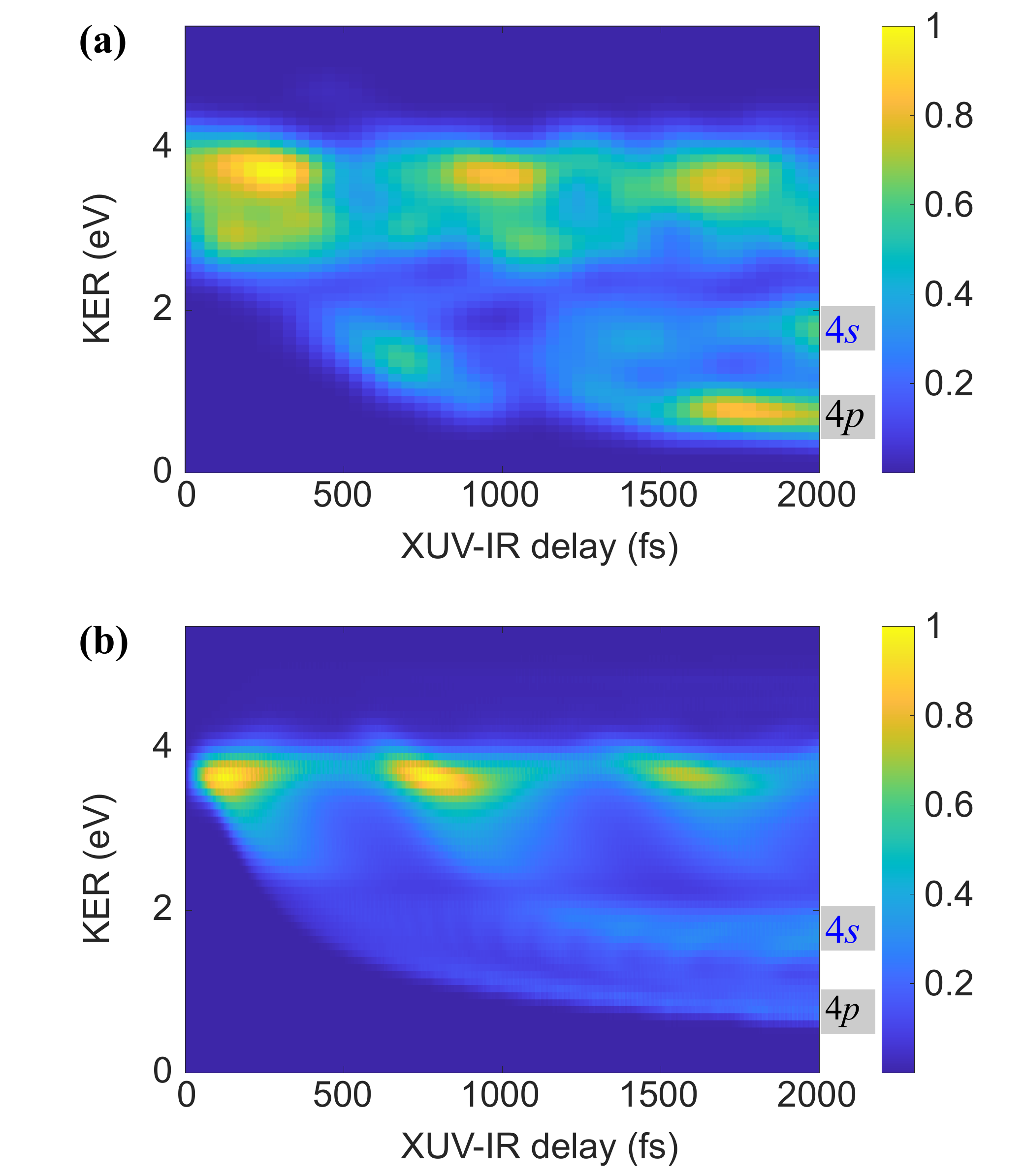}
\end{center}
\caption{Measured \textbf{(a)} and calculated \textbf{(b)} time-resolved KER spectra of the Coulomb explosion channel of Ar$_2$.}
\label{figure2}
\end{figure}

The measured time-resolved KER spectra of the coincident Ar$^{+}$-Ar$^{+}$ channel is shown in Fig.~\ref{figure2}(a). Comparing with the previous study \cite{mizuno2017time}, our result validates that there are two clearly visible decay channels approaching 0.8~eV and 1.9~eV at the largest pump-probe delays, which are identified as $4p$ and $4s$ excited states in the neighboring atoms, respectively. What we observed exclusively is the periodic modulation on the main island at 3.8~eV. The oscillation period is about 800~fs, a typical time scale of vibrational motions of noble-gas dimers. This is reminiscent of the nuclear dynamics in this ultrafast inter-atomic energy transfer process.

\begin{figure}
\begin{center}
\includegraphics[width=8.6cm]{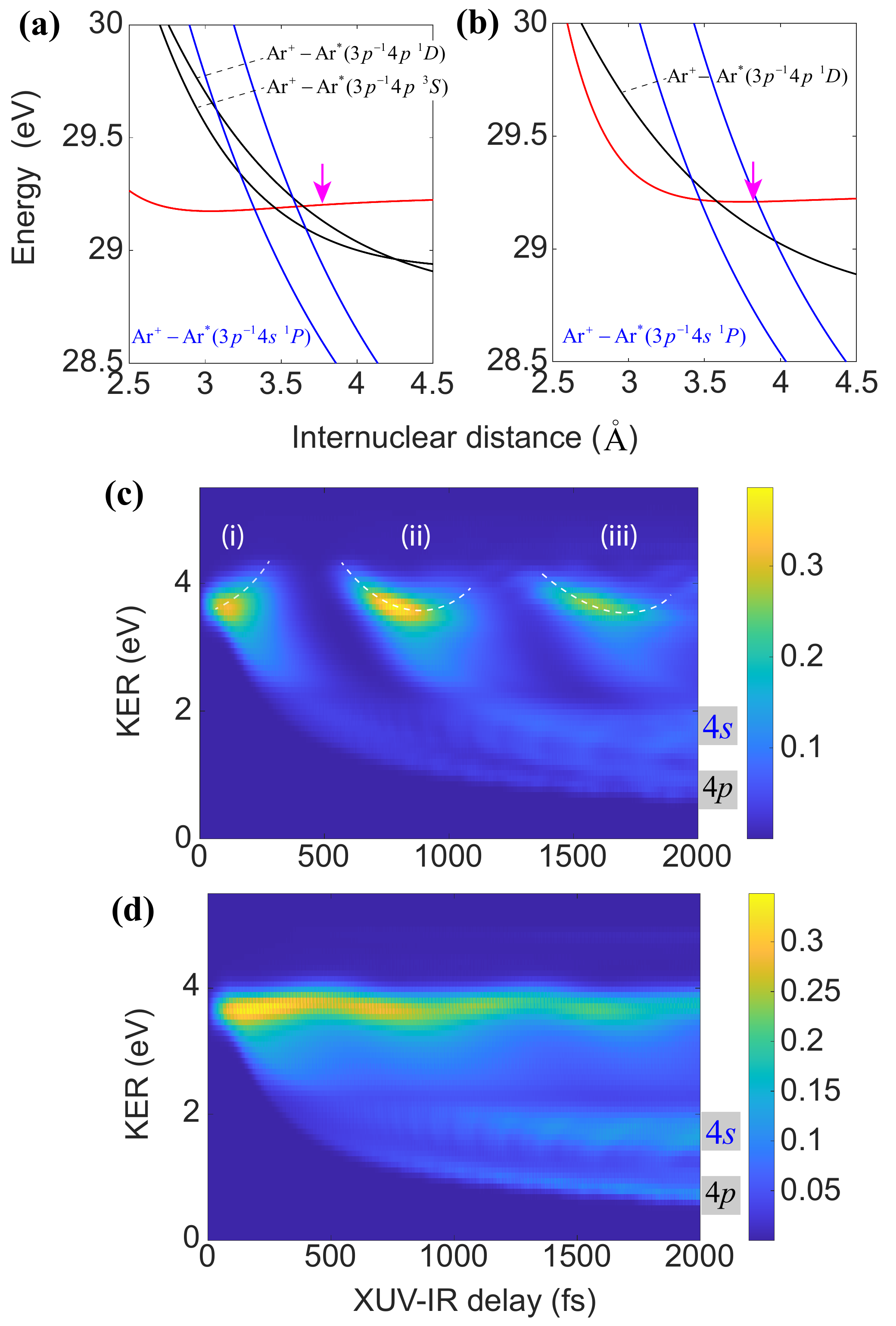}
\end{center}
\caption{Potential-energy curves of argon dimers relevant for the process belonging to ungerade $^2\Sigma_u^{+}$ (\textbf{a}) and gerade $^2\Sigma_g^{+}$ (\textbf{b}) symmetry, taken from Ref.~\cite{mizuno2017time}. The magenta arrows indicate the equilibrium internuclear distance of the ground state of the neutral dimer, i.e. the center of the Franck-Condon region. (\textbf{c-d}) Calculated time-resolved KER spectra from the ungerade and gerade states, respectively.}
\label{figure3}
\end{figure}

To interpret our experimental results, we performed quantum dynamics simulations of the process and computed the time-dependent KER spectra in the framework of nuclear wave packets propagating on the potential-energy curves (PECs) of the involved states. The \textit{ab initio} PECs of the ground and the final Ar$^+(3p^{-1})$-Ar$^+(3p^{-1})$ states are taken from Ref.~\cite{miteva2014interatomic}, while the PECs of the initial and final states for the frustrated ICD process, namely Ar$^+(3s^{-1})$-Ar, and Ar$^+(3p^{-1})$-Ar$^*(3p^{-1}4s)$ and Ar$^+(3p^{-1})$-Ar$^*(3p^{-1}4p)$, are taken from Ref.~\cite{mizuno2017time} and depicted in Figs.~\ref{figure3}~(a) and (b). The complete set of PECs can be found in Ref. \cite{tsveta}. Despite the high-level electronic-structure method used to compute the $3s^{-1}$ PECs (the non-Dyson-ADC(3) Green's function approach \cite{schirmer1998non}), the potential-energy wells appear to be slightly deeper, resulting in a shorter oscillation period than the experimentally observed one. To correct for this, the ungerade \textit{ab initio} $3s^{-1}$ PEC was first fitted to a Morse potential and then its parameters optimized such that the wave-packet motion reproduces the experimentally observed period of about 800~fs. The Morse-potential parameters were ($D_e$=0.129~eV, $R_e$=3.03\AA, $\alpha$=0.91 1/a.u.) from the ab-initio calculations and ($D_e$=0.064~eV, $R_e$=3.03\AA, $\alpha$=0.78 1/a.u.) after optimization. Such an adjustment of PECs of rare-gas dimers has already been successfully used in the past (see, Ref.~\cite{fedyk2023interference}). 

To simulate the energy transfer from Ar$^+(3s^{-1})$-Ar to Ar$^+(3p^{-1})$-Ar$^*(3p^{-1}4s)$ and Ar$^+(3p^{-1})$-Ar$^*(3p^{-1}4p)$, taking place at the points of curve-crossings, we introduced an effective decay width, which corresponds to the energy-transfer time of $\tau_{3s} = 824 \pm 183$~fs extracted in Ref.~\cite{mizuno2017time}. The corresponding decay width of 0.8~meV was statistically distributed to the respective $4s$- and $4p$-excited states according to their multiplicity. The effective decay widths are then associated with each state and introduced with step functions at the corresponding curve-crossings. This approach allows us to use the same coupled set of time-dependent Schrödinger equations to describe the nuclear dynamics as for ICD (see, e.g., Refs.~\cite{scheit2004on,demekhin2009interatomic,demekhin2013overcoming}).

The simulation assumes a broad-band ionization with the XUV-pump pulse, promoting the vibrational ground state of Ar$_2$ to the Ar$^+(3s^{-1})$-Ar PEC. This initial wave packet is then propagated with a Hamiltonian containing all the states involved in the frustrated-ICD process (see, Figs.~\ref{figure3}~(a) and (b)) coupled with the respective decay widths. This approach follows the well-established method to treat electronic decay processes accounting for the nuclear dynamics of the system introduced in Refs.~\cite{cederbaum1993nuclear,Pahl1996competition}. The wave-packet propagation is performed with the complex short iterative Lanczos integrator \cite{Friesner1989method} implemented in the Heidelberg MCTDH package \cite{beck2000mctdh}. To get the KER spectrum at each time step ($\Delta\tau = 5$~fs), the wave packet propagating on the $4s$- or $4p$-excited state is promoted to the final repulsive dicationic state, simulating the ionization by the IR-probe pulse. This procedure is separately repeated for each gerade and ungerade transition between the $3s$-hole and $4s$- or $4p$-excited state. Finally, the gerade and ungerade contributions are incoherently summed-up in order to construct the time-resolved XUV-IR KER spectra shown in Fig.~\ref{figure2}~(b).

\begin{figure}
\begin{center}
\includegraphics[width=5.4cm]{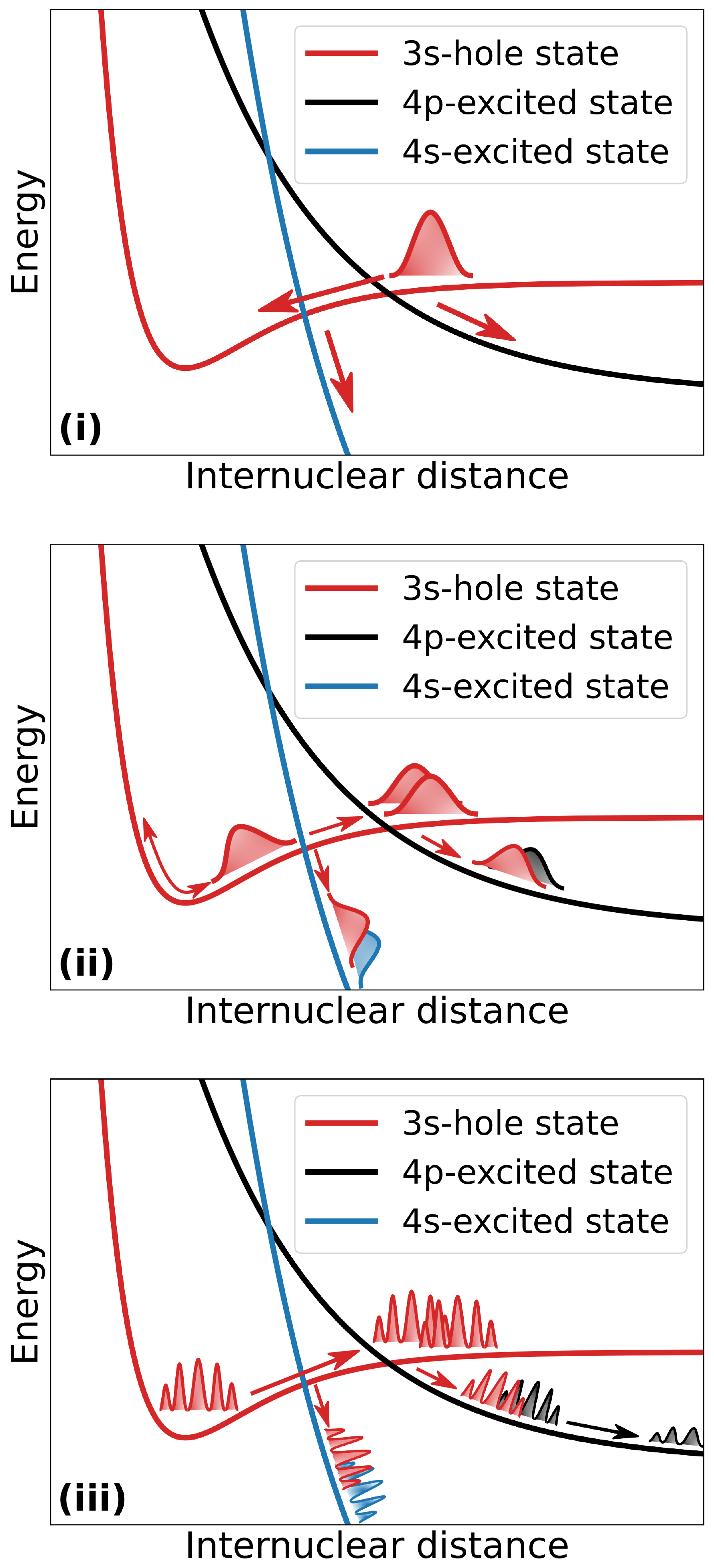}
\end{center}
\caption{(\textbf{i-iii}) Schematic representation of nuclear dynamics in Ar$_2$ during frustrated ICD in the XUV-IR-delay ranges (i), (ii), and (iii), compare Fig.~\ref{figure3}. The red, blue, and black curves represent the $3s$-hole (Ar$^{+*}$(3s$^{-1}$)-Ar), $4s$-, and $4p$-excited (Ar$^{+}$(3p$^{-1}$)-Ar*(3p$^{-1}$ 4s/4p)) states, respectively. The arrows mark the propagation direction.}
\label{figure4}
\end{figure}

As we see from Fig.~\ref{figure2}, the computed time-dependent KER spectra are in good agreement with the experimental one. The structures converging to 0.8~eV and 1.9~eV, associated with the probing of the $4p$- and $4s$-excited state, respectively, the three intense islands at 3.8~eV, resulting from the oscillation of the wave packet in the $3s^{-1}$ potential well, and the lack of signal at 2.3~eV, due, as we will see, to a destructive interference, are well reproduced. This allows us to analyze in detail the quantum dynamics along the frustrated-ICD process and the various features observed in the measured KER spectra. 

Let us start with the slightly parabolic shapes of the three main islands at 3.8~eV. This reflects the motion of the wave packet on the $3s$-hole state and the fact that the curve-crossings with the $4s$- and $4p$-excited states are reached one after the other. The wave packet on the $3s$-hole potential is initially localized around the equilibrium internuclear distance of the neutral dimer (marked by magenta arrows in Fig.~\ref{figure3}~(a) and (b)), and then starts to propagate towards the potential minimum passing through all the curve-crossings, where the population transfers to the $4s$- and $4p$-potentials take place. Because decreasing internuclear distances appear as increasing KER in the spectra, the first island (between 0 and 500~fs) has an increasing parabolic shape. After the wave packet passes the last curve-crossing and moves until its left turning point and back in the $3s^{-1}$ potential well, no frustrated-ICD takes place and correspondingly the KER signal disappears. On its way back, while moving towards increasing internuclear distances, the wave packet passes the curve-crossings in a reversed order, making the KER signal appear as a decreasing parabola up to about 800~fs when first oscillation is completed and the process starts again. This motion of the wave packet and the decay at the respective curve-crossings is schematically depicted in Fig.~\ref{figure4}~(i) and (ii). During its motion, the wave packet expands and develops structure (Fig.~\ref{figure1}~(iii)) making the signal to fade out and blur with increasing pump-probe delay time.

By separating the gerade and ungerade contributions of the spectra, we can get additional insights into nuclear dynamics. Those are shown in Fig.~\ref{figure3}~(c) for the ungerade states, and in Fig.~\ref{figure3}~(d) for the gerade states. The most significant difference between the two is the signal around 3.8~eV, appearing as clear islands in the ungerade case and as a wave-like structure in the gerade spectra. The reason lies in the shape of the $3s^{-1}$ potential and the positions of the curve-crossings. The gerade-symmetry PEC is shallower and the curve-crossings are closer to the potential minimum and the inner turning point (compare Fig.~\ref{figure3}~(a) and (b)). Consequently, the oscillation period for the gerade contribution is slightly longer and as some portion of the wave packet is always at the curve crossings, the signal at 3.8~eV does not disappear. The wave-like shape, however, clearly reflects the back-and-forth motion of the wave packet. 

We now return to the lack of signal around 2.3~eV, appearing in both gerade and ungereade spectra, and clearly visible also in the experimental KER-trace (see Fig.~\ref{figure2}~(a)). Our analysis shows that this is a result of a destructive interference taking place between different portions of the wave packet transferred to the $4s$- and $4p$-excited potentials at two consecutive passages through the corresponding curve-crossings. In other words, during its excursion to the inner turning point and back, the wave packet moving on the $3s^{-1}$ PEC accumulates a phase that is close to an odd multiple of $\pi$ and when transferred to $4s$ and $4p$ potentials interferes destructively with the portions of the wave packet already propagating on these PECs (see Fig.~\ref{figure4}~(ii) and (iii)). 


Before we conclude, we would like to briefly comment on one feature in the measured KER spectra that is not reproduced by the simulations. This is the decrease in the signal between 800~fs and 1200~fs of the trace converging to 0.8~eV and reflecting the dynamics on the PEC of the $4p$-excited state. A possible explanation of this feature could be that around 1~ps after the energy transfer, the wave packet propagating on the $4s$ state can be efficiently transferred to a higher excited Ar$^+(3p^{-1})$-Ar$^*(3p^{-1}nl)$ state by a single IR photon, thus removing the population from the measured Ar$^+$-Ar$^+$ signal. Indeed, the ionization of the $4p$-electron needs between 2 and 3 photons, depending on the internuclear distance, and a resonance condition with a single photon will easily hinder the multi-photon ionization channel. A possible candidate will be Ar$^+(3p^{-1})$-Ar$^*(3p^{-1}4d)$ state which would satisfy the resonant conditions, but the spectra of excited states in this energy range is already so dense that a definitive assignment of such transitions becomes challenging.    

In summary, we have demonstrated that the high-resolution time-resolved photoion-photoion spectroscopy performed with high-harmonic sources can give access to the quantum nuclear dynamics underlying excitation-energy transfer, i.e. frustrated ICD, in argon dimers. The different features in the time-resolved KER spectra can be analyzed with the help of quantum wave-packet propagation simulations, leading to a detailed reconstruction of the nuclear dynamics throughout the process. The vibrational motion of the system in the initially populated $3s^{-1}$ state, as well as the interference of different portions of the wave packet transferred to the final $4s$- and $4p$-excited states at different times, lead to characteristic patterns in the time-resolved KER spectrogram. The high level of details that can be achieved when analyzing the frustrated-ICD process in rare-gas dimers suggests that this technique can be used to study such energy-transfer processes in more complex systems like weakly-bound molecular dimers, as well as (micro-)solvated molecules. We hope that our study will stimulate further efforts in this direction. 

\vspace{0.2cm}

M. Han and J. Fedyk contributed equally to this work. We thank K. Ueda for fruitful discussions. We thank A. Schneider and M. Seiler for their technical support in performing the experiment, and T. Miteva for providing the \textit{ab initio} potential-energy curves of Ar$_2$. M. Han acknowledges funding from the European Union’s Horizon 2020 research and innovation programme under Marie Skłodowska-Curie grant agreement no. 801459, FP-RESOMUS. J. Fedyk and A. I. Kuleff acknowledge the financial support by the European Research Council (ERC) (Advanced Investigator Grant No. 692657).

\bibliography{ref}

\end{document}